\documentclass[prd,twocolumn,nofootinbib]{revtex4}

\usepackage{amsmath,amssymb}
\usepackage{mathrsfs}
\usepackage{mathtools}
\usepackage{rotating}
\usepackage{mathrsfs}
\usepackage{pstricks}
\usepackage{pst-func}
\usepackage{graphicx}
\usepackage{cases}

\def\BibTeX{{\rm B\kern-.05em{\sc i\kern-.025em b}\kern-.08em
    T\kern-.1667em\lower.7ex\hbox{E}\kern-.125emX}}

\begin{document}

\title{Reply to ``Comment on `Bell's Theorem Versus Local Realism in a Quaternionic Model of Physical Space'''}

\author{Joy Christian}

\email{jjc@bu.edu}

\affiliation{Einstein Centre for Local-Realistic Physics, 15 Thackley End, Oxford OX2 6LB, United Kingdom}

\begin{abstract}
In this paper, I respond to a critique of one of my papers previously published in this journal, entitled ``Bell's Theorem Versus Local Realism in a Quaternionic Model of Physical Space.'' That paper presents a local-realistic model of quantum correlations based on a quaternionic 3-sphere, taken as a physical space in which we are confined to perform all our experiments. The critique, on the other hand, considers two entirely different models within a flat Euclidean space, neither related to my quaternionic 3-sphere model. It then criticizes its own flat space models and claims that it has thereby criticized the model presented in my paper. Along the way, without providing evidence or proof, it claims that the results in my paper are based on mistakes. I demonstrate that there are no mistakes of any kind in my paper. On the contrary, I bring out a number of elementary mathematical and conceptual mistakes from the critique and the critiques it relies on.
\end{abstract}

\maketitle

\section{Introduction}\label{intro}

This reply paper should be read as a continuation of my previous reply paper \cite{IEEE-3}, which is a reply published in this journal to a previous critique of one of my papers by the same author \cite{Gill-IEEE-1}. I have also responded to earlier\break critiques by the same author elsewhere \cite{Reply-Gill,Reply-Gill-IJTP,Scott}. It is unfortunate that the critique in \cite{Gill-IEEE-2} has repeated many arguments that I have already addressed in \cite{IEEE-3,Reply-Gill,Reply-Gill-IJTP,Scott}, as well as in Appendix~B of \cite{IEEE-1} and Appendix~B of \cite{IEEE-2}. Sections I, II, and III of my previous reply \cite{IEEE-3} provide the background for understanding the current reply to the critique in \cite{Gill-IEEE-2}. In what follows, I have tried to avoid repetitions of my refutations of the issues raised in \cite{Gill-IEEE-2} that are already addressed by me in \cite{IEEE-3}. Needless to say, prior reading of my original paper \cite{IEEE-1} and its pedagogical exposition \cite{IEEE-2} is recommended for understanding this reply.

The central result I have presented in \cite{IEEE-1} and \cite{IEEE-2}, as well as in my earlier papers \cite{Disproof,IJTP,RSOS}, is this special theorem:

{\it Theorem 1:} Quantum mechanical correlations predicted by the entangled singlet state can be understood as classical, local, realistic, and deterministic correlations among the pairs of limiting scalar points of a quaternionic 3-sphere, or $S^3$, taken as a model of the three-dimensional physical space.

I have proved this theorem in \cite{IEEE-1,IEEE-2,Disproof,IJTP,RSOS} in several different ways. In fact, its proof can be presented on a single page \cite{disproof}. Its more complete proof can be found in Section III of \cite{IEEE-3}.

Now $S^3$ turns out to be sufficient for understanding the singlet correlations local-realistically, but not sufficient for understanding more general correlations. What is required is an algebraic representation space of $S^3$, which turns out to be an octonion-like 7-sphere, leading to this general theorem:

{\it Theorem 2:} Quantum mechanical correlations predicted by any arbitrary quantum state can be understood as classical, local, realistic, and deterministic correlations among the pairs of limiting scalar points of an octonion-like 7-sphere, as an algebraic representation space of the quaternionic 3-sphere.

I have proved this general theorem in \cite{RSOS}. For a summary, see also \cite{slides}. According to these two theorems, the geometrical and topological properties of the quaternionic 3-sphere, or $S^3$, are the {\it raison d'\^etre} of the origins and strengths of all quantum correlations, not quantum entanglement {\it per se}. The latter is merely a placeholder for quantum correlations, similar to how phlogiston was for combustion until Lavoisier proved it to be a rapid oxidation, and action-at-a-distance was for Newton's gravity until Einstein recognized it to be an effect of the curvature of spacetime. Tsirel’son’s bounds
\begin{equation}
-2\sqrt{2}\leqslant{\cal E}({\mathbf{a}},{\mathbf{b}})+{\cal E}({\mathbf{a}},{\mathbf{b'}})+{\cal E}({\mathbf{a'}},{\mathbf{b}})-{\cal E}({\mathbf{a'}},{\mathbf{b'}})\leqslant 2\sqrt{2}
\end{equation}
turn out to be consequences of the above hypothesis \cite{IEEE-1,RSOS}. Although in my view the strong correlations observed in Nature is evidence enough in favor of this hypothesis, in \cite{IJTP} I have proposed an experiment set in a macroscopic domain, devoid of any quantum features, that may be able to falsify it.

In this paper, I demonstrate, point by point, that none of the claims made in the critique \cite{Gill-IEEE-2}, or in the critiques mentioned therein, undermine the above theorems. Contrary to its claims, the critique has not found any kind of mistakes in my papers, or in the numerical simulations of the quaternionic 3-sphere model cited therein. On the contrary, in what follows I point out a number of mathematical and conceptual mistakes in the critique \cite{Gill-IEEE-2} and the critiques \cite{Gill-IEEE-1,Gill-Entropy,Lasenby-AACA} it relies on.

The common defect in the critiques \cite{Gill-IEEE-1}, \cite{Gill-IEEE-2}, and \cite{Gill-Entropy} is that they consider and criticize models that are entirely different from the one presented in \cite{IEEE-1} and \cite{IEEE-2}, and then claim that they have thereby criticized the model presented in my papers. This has been an effective strategy for the critiques in \cite{Gill-IEEE-1}, \cite{Gill-IEEE-2}, and \cite{Gill-Entropy} despite its manifest logical fallacy. For example, it can be easily verified by searching the PDF file of the critique \cite{Gill-IEEE-2} that in its text the word ``quaternion'' appears only once, and that only in a dismissive sentence: "But Christian’s speculations about a quaternionic space seem completely irrelevant." And the word ``3-sphere'' also appears only once, and that only within a sentence quoted from {\it my} simulation code \cite{rpubs}. The critique thus entirely avoids engaging with the details of the quaternionic 3-sphere model, overlooking many of its vital features and their broader physical significance.

The minimum prerequisite for understanding the model of quantum correlations presented in \cite{IEEE-1} and \cite{IEEE-2} (as well as in \cite{Disproof,IJTP,RSOS}) is at least some appreciation that it is based on a quaternionic 3-sphere, or $S^3$, hypothesized to be the physical space in which we are inescapably confined to perform our Bell-test experiments. The origins and strengths of quantum correlations are then {\it inevitable} consequences of the Clifford algebra and geometry of this physical space, as we will see in the next Section for the singlet state. By contrast, the flatland or ${\mathrm{I\!R}^3}$ perspective and the use of vector algebra adhered to in the critiques \cite{Gill-IEEE-1,Gill-IEEE-2} leads one to the traditional interpretation of Bell's theorem, as I have demonstrated in Section X of \cite{IEEE-1}.

\section{R\'esum\'e of the quaternionic 3-sphere model} \label{II}

Within the quaternionic 3-sphere model presented in \cite{IEEE-1,IEEE-2}, the important question to be answered is this: What will be the value of the joint result ${{\mathscr A}{\mathscr B}({\mathbf a},{\mathbf b},\lambda)=\pm1}$ when the individual results ${\mathscr A}({\mathbf a},\lambda)=\pm1$ and ${\mathscr B}({\mathbf b},\lambda)=\pm1$ are observed by Alice and Bob separately but simultaneously, in ``coincidence counts'', {\it within} a space-like hypersurface $S^3$ within spacetime? Here ${\mathbf a}$ and ${\mathbf b}$ are the detector directions, chosen by Alice and Bob, and $\lambda$ is a common cause, so that
\begin{equation}
{\mathscr A}{\mathscr B}({\mathbf a},{\mathbf b},\lambda)={\mathscr A}({\mathbf a},\lambda)\,{\mathscr B}({\mathbf b},\lambda), \label{fac}
\end{equation}
which is the factorizability condition for the results required by local causality. To answer the above question, recall that there are three different sets of experiments involved in any Bell-test experiment \cite{IEEE-3}. In the first two sets of experiments Alice and Bob make spin measurements at their respective stations obtaining the results $\pm1$, with 50/50 chance, so that both $\langle {\mathscr A} \rangle = 0$ and $\langle {\mathscr B} \rangle = 0$. These are two independent sets of experiments in which Alice performs her experiments regardless of Bob’s existence, and vice versa. Their results, ${\mathscr A} = \pm1$, $\langle {\mathscr A} \rangle = 0$ and ${\mathscr B} = \pm1$, $\langle {\mathscr B} \rangle = 0$, are what the measurement functions (66) and (67) defined in \cite{IEEE-1} predict. They describe the detection processes of Alice and Bob as limiting scalar points of two separate quaternions within $S^3$:
\begin{align}
S^3\ni{\mathscr A}&({\mathbf a},{\lambda})\,=\lim_{{\mathbf s}_1\,\rightarrow\,{\mathbf a}}\left\{-\,{\mathbf D}({\mathbf a})\,{\mathbf L}({\mathbf s}_1,\,\lambda)\right\} \\
&=\lim_{{\mathbf s}_1\,\rightarrow\,{\mathbf a}}\left\{\,+\,{\mathbf q}(\eta_{{\mathbf a}{\mathbf s}_1},\,{\mathbf r}_1)\right\} \label{a-q} \\
&=+\lambda\lim_{{\mathbf s}_1\,\rightarrow\,{\mathbf a}}\left\{ \cos(\eta_{{\mathbf a}{\mathbf s}_1})+(I\cdot{\mathbf r}_{1})\sin(\eta_{{\mathbf a}{\mathbf s}_1})\right\}
\end{align}
implying that, as the angle $\eta_{{\mathbf a}{\mathbf s}_1}\rightarrow\,0$,
\begin{equation}
{\mathscr A}({\mathbf a},{\lambda})\rightarrow\pm1 \;\;\;\text{with}\;\;\;\langle\,{\mathscr A}({\mathbf a},{\lambda})\rangle=0, \label{7no}
\end{equation}
and
\begin{align}
S^3\ni{\mathscr B}&({\mathbf b},\,{\lambda})\,=\lim_{{\mathbf s}_2\,\rightarrow\,{\mathbf b}}\left\{+\,{\mathbf L}({\mathbf s}_2,\,\lambda)\,{\mathbf D}({\mathbf b})\right\} \\
&=\lim_{{\mathbf s}_2\,\rightarrow\,{\mathbf b}}\left\{\,-\,{\mathbf q}(\eta_{{\mathbf s}_2{\mathbf b}},\,{\mathbf r}_2)\right\} \label{b-q} \\
&=-\lambda\lim_{{\mathbf s}_2\,\rightarrow\,{\mathbf b}}\left\{ \cos(\eta_{{\mathbf s}_2{\mathbf b}})+(I\cdot{\mathbf r}_{2})\sin(\eta_{{\mathbf s}_2{\mathbf b}})\right\}
\end{align}
implying that, as the angle $\eta_{{\mathbf s}_2{\mathbf b}}\rightarrow\,0$,
\begin{equation}
{\mathscr B}({\mathbf b},{\lambda})\longrightarrow\mp1 \;\;\;\text{with}\;\;\;\langle\,{\mathscr B}({\mathbf b},{\lambda})\rangle=0, \label{8no}
\end{equation}
where ${\mathbf L}({\mathbf s}_1,\,\lambda)$ and ${\mathbf L}({\mathbf s}_2,\,\lambda)$ are spin bivectors and ${\mathbf D}({\mathbf a})$ and ${\mathbf D}({\mathbf b})$ are detector bivectors of Alice and Bob, whose orientations are defined {\it relative} to those of the spin bivectors,
\begin{equation}
{\mathbf L}({\mathbf n},\,\lambda)\,=\,\lambda\,{\mathbf D}({\mathbf n})\,\,\Longleftrightarrow\,\,{\mathbf D}({\mathbf n})\,=\,\lambda\,{\mathbf L}({\mathbf n},\,\lambda)\,, \label{20no}
\end{equation}
$\eta_{{\mathbf a}{\mathbf s}_1}$ and $\eta_{{\mathbf s}_2{\mathbf b}}$ are angles between the detectors and spins, and
\begin{equation}
{\mathbf r}_1=\frac{{\mathbf a}\times{\mathbf s}_1}{||{\mathbf a}\times{\mathbf s}_1||}\;\;\;\text{and}\;\;\;{\mathbf r}_2=\frac{{\mathbf s}_2\times{\mathbf b}}{||{\mathbf s}_2\times{\mathbf b}||}.
\end{equation}

On the other hand, the above equations (\ref{7no}) and (\ref{8no}) (or, equivalently, equations (66) and (67) of \cite{IEEE-1}) do not tell us anything about the third set of experiments carried out by Alice and Bob in which they jointly and simultaneously make measurements within the quaternionic 3-sphere, taken as the physical space. This is because the geometrical properties of the quaternionic 3-sphere are highly nontrivial and they must be respected to respect the hypothesis of the model. Unlike the $\mathrm{I\!R}^3$ model of the physical space, $S^3$ is a compact space that remains closed under multiplications of the quaternions that constitute it. It is defined as a set of all unit quaternions,
\begin{equation}
S^3:=\left\{\,{\mathbf q}(\psi,\,{\mathbf r}):=\exp\left[\,{\mathbf J}({\mathbf r})\,\frac{\psi}{2}\,\right]
\Bigg|\;||\,{\mathbf q}(\psi,\,{\mathbf r})\,||^2=1\right\}\!, \label{nonsin}
\end{equation}
where ${{\mathbf J}({\mathbf r})}$ is a bivector (or pure quaternion) rotating about ${{\mathbf r}\in{\mathrm{I\!R}}^3}$ with the rotation angle ${\psi}$ in the range ${0\leq\psi < 4\pi}$. Here the conventions of Geometric Algebra are used \cite{Clifford}. 

The correct values of the joint result ${{\mathscr A}{\mathscr B}({\mathbf a},{\mathbf b},\lambda)}$ can thus be inferred only by respecting the geometrical properties of $S^3$. They can be preserved only by following the correct rules of quaternionic multiplication. But since the above concepts and the next steps they lead to are not well understood in the critiques \cite{Gill-IEEE-1,Gill-IEEE-2,Gill-Entropy}, let me explain them here step by step. 

\begin{figure}[t]
\centering
\includegraphics[scale=0.8]{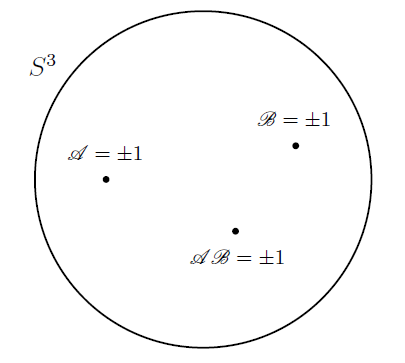}
\caption{The results ${\mathscr A}$ and ${\mathscr B}$ are scalar points of a quaternionic 3-sphere, or ${S^3}$. Since ${S^3}$ remains closed under multiplication, the product ${{\mathscr A}{\mathscr B}}$ is also a point of ${S^3}$, with its binary value ${\pm1}$ dictated by the geometry of $S^3$. After \cite{IEEE-3}.}
\label{Fig-1}
\end{figure}

The crucial question is: What will be the value of the joint result ${{\mathscr A}{\mathscr B}({\mathbf a},{\mathbf b},\lambda)}$ within $S^3$? Can it simply be the product of the values seen in (6) and (10)? No, that cannot be a meaningful value within $S^3$. The values seen in (6) and (10) are scalar numbers, not quaternions. Scalar numbers do not respect the multiplication rules of quaternions. But, by definition, $S^3$ in (12), taken as the physical space, is the set of all unit quaternions. Therefore the value of the joint result ${{\mathscr A}{\mathscr B}({\mathbf a},{\mathbf b},\lambda)}$ must be a limiting scalar point of a quaternion within $S^3$ as depicted in Fig.~\ref{Fig-1}, just as the individual results ${{\mathscr A}({\mathbf a},\lambda)}$ and
${{\mathscr B}({\mathbf b},\lambda)}$ are the limiting scalar points of two quaternions, as specified in the equations (\ref{a-q}) and (\ref{b-q}) above.

\begin{figure}[t]
\vspace{-0.25cm}
\centering
\includegraphics[scale=0.56]{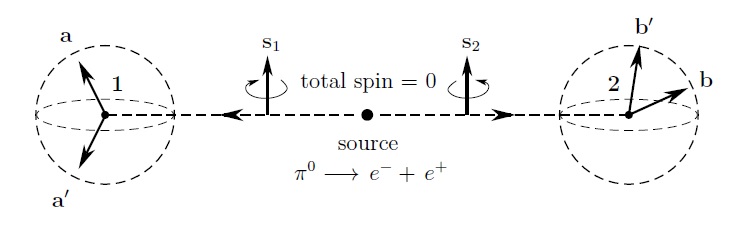}
\caption{A spin-less neutral pion decays into an electron-positron pair. Measurements of spin components on each separated fermion are performed by Alice and Bob at remote stations ${\mathbf{1}}$ and ${\mathbf{2}}$, providing binary outcomes along freely chosen directions ${\mathbf a}$ and ${\mathbf b}$. The conservation of spin momentum dictates that the net spin of the pair remains zero during the free evolution. After \cite{IEEE-1}.}
\vspace{0.2cm}
\label{Fig-2n}
\end{figure}

The next question is: Which quaternion within $S^3$ would give the correct value of the joint result ${{\mathscr A}{\mathscr B}({\mathbf a},{\mathbf b},\lambda)=\pm1}$ as its limiting scalar point while the results ${{\mathscr A}({\mathbf a},\lambda)}$ and ${{\mathscr B}({\mathbf b},\lambda)}$ are observed by Alice and Bob simultaneously in coincidence counts? The answer to this question can be found from the condition (\ref{fac}) and the definitions (\ref{a-q}) and (\ref{b-q}) of the individual results ${{\mathscr A}({\mathbf a},\lambda)}$ and ${{\mathscr B}({\mathbf b},\lambda)}$ within $S^3$, giving
\begin{align}
S^3\ni\;&{\mathscr A}{\mathscr B}({\mathbf a},\,{\mathbf b},\,{\lambda})={\mathscr A}({\mathbf a},\,{\lambda})\,{\mathscr B}({\mathbf b},\,{\lambda}) \\
&\!=\!\left[\lim_{{\mathbf s}_1\,\rightarrow\,{\mathbf a}}\left\{\,+\,{\mathbf q}(\eta_{{\mathbf a}{\mathbf s}_1},\,{\mathbf r}_1)\right\}\right]\!\left[\lim_{{\mathbf s}_2\,\rightarrow\,{\mathbf b}}\left\{\,-\,{\mathbf q}(\eta_{{\mathbf s}_2{\mathbf b}},\,{\mathbf r}_2)\right\}\right]\!.
\end{align}
The question now is: How must we evaluate the last product without compromising the properties of $S^3$ that requires us to respect the quaternionic rules of multiplication when the results ${{\mathscr A}({\mathbf a},\lambda)}$ and ${{\mathscr B}({\mathbf b},\lambda)}$ are observed simultaneously? If the conservation  of zero spin angular momentum is satisfied, as it has to be, which requires us to set ${\mathbf s}_1 = {\mathbf s}_2$ (cf. Fig.~\ref{Fig-2n}), then the correct answer to the above question is: By using the ``product of limits equal to the limits of product'' rule, giving
\begin{align}
S^3\ni\;&{\mathscr A}({\mathbf a},\,{\lambda})\,{\mathscr B}({\mathbf b},\,{\lambda}) \notag \\
&\!=\left[\lim_{{\mathbf s}_1\,\rightarrow\,{\mathbf a}}\left\{\,+\,{\mathbf q}(\eta_{{\mathbf a}{\mathbf s}_1},\,{\mathbf r}_1)\right\}\right]\!\left[\lim_{{\mathbf s}_2\,\rightarrow\,{\mathbf b}}\left\{\,-\,{\mathbf q}(\eta_{{\mathbf s}_2{\mathbf b}},\,{\mathbf r}_2)\right\}\right] \\
&\!=\lim_{\substack{{\mathbf s}_1\,\rightarrow\,{\mathbf a} \\ {\mathbf s}_2\,\rightarrow\,{\mathbf b}}}\left\{\,-\,{\mathbf q}(\eta_{{\mathbf a}{\mathbf s}_1},\,{\mathbf r}_1)\,{\mathbf q}(\eta_{{\mathbf s}_2{\mathbf b}},\,{\mathbf r}_2)\right\} \label{11} \\
&\!=\lim_{\substack{{\mathbf s}_1\,\rightarrow\,{\mathbf a} \\ {\mathbf s}_2\,\rightarrow\,{\mathbf b}}}\left\{\,-\,{\mathbf q}(\eta_{{\mathbf u}{\mathbf v}},\,{\mathbf r}_{0})\right\},
\end{align}
where ${\mathbf q}(\eta_{{\mathbf u}{\mathbf v}},\,{\mathbf r}_{0})={\mathbf q}(\eta_{{\mathbf a}{\mathbf s}_1},\,{\mathbf r}_{1})\,{\mathbf q}(\eta_{{\mathbf s}_2{\mathbf b}},\,{\mathbf r}_{2})$ is necessarily another unit quaternion within $S^3$ because the latter remains closed under multiplication, with half of its rotation angle 
\begin{align}
\eta_{{\mathbf u}{\mathbf v}}=\cos^{-1}&\big\{({\mathbf a}\cdot{\mathbf s}_1)({\mathbf s}_2\cdot{\mathbf b}) \notag \\
&\;\;\;-({\mathbf a}\cdot{\mathbf s}_2)({\mathbf s}_1\cdot{\mathbf b})+({\mathbf a}\cdot{\mathbf b})({\mathbf s}_1\cdot{\mathbf s}_2)\big\} \label{38}
\end{align}
and the axis of its rotation
\begin{align}
{\mathbf r}_{0}&=\frac{({\mathbf a}\cdot{\mathbf s}_1)({\mathbf s}_2\!\times\!{\mathbf b})\!+\!({\mathbf s}_2\cdot{\mathbf b})({\mathbf a}\!\times\!{\mathbf s}_1)\!-\!({\mathbf a}\!\times\!{\mathbf s}_1)\!\times\!({\mathbf s}_2\!\times\!{\mathbf b})}{\sin\left(\,\eta_{{\mathbf u}{\mathbf v}}\right)}.
\end{align}
Note, however, that the above results do not depend on using the ``product of limits equal to the limits of product'' rule, as long as the conservation of angular momentum is respected. 

Now the conservation of spin angular momentum requires ${\mathbf s}_1 = {\mathbf s}_2$ (cf. Fig.~\ref{Fig-2n}).
And from (\ref{38}) we note that for ${\mathbf s}_1 = {\mathbf s}_2$ the half angle $\eta_{{\mathbf u}{\mathbf v}}$ reduces to $\eta_{{\mathbf a}{\mathbf b}}$, reducing ${\mathbf q}(\eta_{{\mathbf u}{\mathbf v}},\,{\mathbf r}_{0})$ to 
\begin{equation}
{\mathbf q}(\eta_{{\mathbf a}{\mathbf b}},\,{\mathbf r}_{0})=\cos(\eta_{{\mathbf a}{\mathbf b}}) + (I\cdot{\mathbf r}_0)\,\sin(\eta_{{\mathbf a}{\mathbf b}}).
\end{equation}
On the other hand, regardless of whether ${\mathbf s}_1 = {\mathbf s}_2$ or  ${\mathbf s}_1 \not= {\mathbf s}_2$,
\begin{equation}
\lim_{\substack{{\mathbf s}_1\,\rightarrow\,{\mathbf a} \\ {\mathbf s}_2\,\rightarrow\,{\mathbf b}}}{\mathbf r}_{0}({\mathbf s}_1,\,{\mathbf s}_2)=\vec{\,\mathbf 0} \label{21-n}
\end{equation}
during the detection process of the joint result ${\mathscr A}{\mathscr B}({\mathbf a},\,{\mathbf b},\,{\lambda})$. The value ${\mathscr A}{\mathscr B}$ of the product 
${\mathscr A}({\mathbf a},\lambda)\,{\mathscr B}({\mathbf b},\lambda)$ thus tends to
\begin{align}
{\mathscr A}&({\mathbf a},\lambda)\,{\mathscr B}({\mathbf b},\lambda)={\mathscr A}{\mathscr B}({\mathbf a},\,{\mathbf b},\,{\lambda}) \notag \\
&=\!\lim_{\substack{{\mathbf s}_1\,\rightarrow\,{\mathbf a} \\ {\mathbf s}_2\,\rightarrow\,{\mathbf b}}}\left\{\,-\,{\mathbf q}(\eta_{{\mathbf a}{\mathbf b}},\,{\mathbf r}_{0})\right\}\rightarrow
\begin{cases}
\!\!-1 &\!\!\text{if}\;\,{\mathbf s}_1\not={\mathbf s}_2 \\
\!\!-\,{\mathbf a}\cdot{\mathbf b} &\!\!\text{if}\;\,{\mathbf s}_1={\mathbf s}_2.
\end{cases} \label{74-n}
\end{align}
Given the equality ${\mathbf s}_1={\mathbf s}_2$ required by the conservation of zero spin angular momentum (cf. Fig.~\ref{Fig-2n}), this tendency (\ref{74-n}) for the value ${\mathscr A}{\mathscr B}$ holds for each run of the experiment. As a result, the correlation between the results ${{\mathscr A}({\mathbf a},\,{\lambda})}$ and ${{\mathscr B}({\mathbf b},\,{\lambda})}$ observed by Alice and Bob within $S^3$ works out as follows. Since $\lambda$ is a fair coin, we can assume uniform averaging over $\lambda$ with probability density $p(\lambda)=\frac{1}{n}$, where $n$ is the total number of experimental trials. This allows us to reduce the expectation function ${\cal E}({\mathbf a},\,{\mathbf b})$ to a discrete version:
\begin{align}
{\cal E}_{\mathrm{L.R.}}({\mathbf a},\,{\mathbf b})
&=\!\int
{\mathscr A}({\mathbf a},\,\lambda)\,{\mathscr B}({\mathbf b},\,\lambda)\;p(\lambda)\,d\lambda \\
&\approx\!\lim_{\,n\,\gg\,1}\left[\frac{1}{n}\sum_{k\,=\,1}^{n}\,{\mathscr A}({\mathbf a},\,{\lambda}^k)\;{\mathscr B}({\mathbf b},\,{\lambda}^k)\right] \label{57a} \\
&=\lim_{\,n\,\gg\,1}\left[\frac{1}{n}\sum_{k\,=\,1}^{n}\,\lim_{\substack{{\mathbf s}_1\,\rightarrow\,{\mathbf a} \\ {\mathbf s}_2\,\rightarrow\,{\mathbf b}}}\left\{\,-\,{\mathbf q}(\eta_{{\mathbf a}{\mathbf b}},\,{\mathbf r}_{0})\right\}\right] \label{60a}\\
&=\,-\cos(\,\eta_{{\mathbf a}{\mathbf b}}). \label{65a}
\end{align}
This corroborates the hypothesis put forward in \cite{Disproof,IJTP,RSOS,IEEE-1,IEEE-2} that the observed singlet correlations are correlations among the limiting scalar points ${{\mathscr A}({\mathbf a},\,{\lambda})=\pm1}$ and ${{\mathscr B}({\mathbf b},\,{\lambda})=\pm1}$ of a quaternionic 3-sphere. I have derived the correlation (\ref{65a}) in 
\cite{Disproof,IJTP,RSOS,IEEE-1,IEEE-2} in several different ways, proving the {\it Theorem~1} stated above in the Introduction. See also its derivation in equations (34) to (41e) of Section III in my previous reply~\cite{IEEE-3}.

It is evident from the above derivation that the quaternionic 3-sphere model is not a model of after-the-events procedure followed by experimenters within ${\mathrm{I\!R}^3}$ as implicitly assumed in the critique, but a {\it theoretical} model that predicts the values of the simultaneous events ${\mathscr A}$, ${\mathscr B}$, and ${{\mathscr A}{\mathscr B}}$ occurring within $S^3$, dictated by its geometrical and topological properties. Thus the critique \cite{Gill-IEEE-2} confuses epistemology with ontology, committing what philosophers call ``a category error.''

\section{Point by point response to the critique}

With the above summary of the 3-sphere model, let us now proceed to address the issues raised in the critique \cite{Gill-IEEE-2}. While there are many incorrect and unproven claims throughout the critique, I focus in this Section on those that are significant.

\subsection{Concerning the status of Bell's theorem}

In its Introduction, the critique in \cite{Gill-IEEE-2} begins with a serious misconception that “Bell’s Theorem” is a proven theorem in the mathematical sense and therefore any critique of it must be flawed. But while even proven mathematical theorems may not be immune to refutations by counterexamples as so lucidly explained by Lakatos \cite{Lakatos}, Bell's theorem is not a theorem in the mathematical sense to begin with, as I have pointed out in Answer~1 of Appendix~B in \cite{IEEE-1}. It is a {\it physical} argument, based on a number of implicit and explicit physical assumptions, which can be and have been questioned before, not only by me \cite{IEEE-1,IEEE-2,Disproof,IJTP,RSOS} but also by many others (cf. footnote~1 in \cite{IEEE-1}). If it were a proven theorem in the mathematical sense, then it would not require physical experiments for its validity and any loophole (or "gap") would render it invalid. 

Moreover, as I have discussed in detail in Section~II of \cite{IEEE-3}, the mathematical part of Bell’s theorem is nothing more than a reworking of an inequality in probability theory proven by Boole one hundred and eleven years before the publication in 1964 of Bell's physical argument based on it \cite{Boole-1,Boole-2,Bell-1964}. The critique in \cite{Gill-IEEE-2} seems to be aware of the historical precedence of Boole's mathematical inequality, but fails to appreciate the distinction between a purely mathematical inequality and the radical metaphysical claims regarding the nature of physical reality made by Bell based on that mathematical inequality. This is unfortunate, because it is quite well known that Boole's inequality can be derived without assuming locality, as, for example, I have derived in Section~4.2 of \cite{RSOS}, or without assuming realism, as, for example, Eberhard has derived in \cite{Eberhard}, or without assuming either locality or realism, as, for example, Boole has derived in \cite{Boole-1,Boole-2}. By now it is well known that, when viewed as a mathematical theorem about a probability distribution and its marginals, the theorem part of Bell's argument ({\it i.e.,} Boole's inequality) can be derived without any reference to local realism at all.

Bell’s argument, on the other hand, starts off by assuming local realism. It then proceeds to re-derive Boole's inequality without referring to Boole. Since this inequality is not consistent with quantum mechanical probabilities, Bell argues that quantum mechanics is inconsistent with local realism. But given the different derivations of Boole's inequality in Section~4.2 of \cite{RSOS}, \cite{Eberhard}, and \cite{Boole-1,Boole-2} mentioned in the previous paragraph, we are equally justified in concluding that quantum mechanics is inconsistent with non-local realism, or local non-realism, or non-local non-realism, respectively. Indeed, one can start off by explicitly assuming non-locality and non-realism and still derive Boole's inequality, requiring only the additivity of expectation values \cite{Oversight} and that results of incompatible experiments along mutually exclusive measurement directions can occur in nature simultaneously \cite{RSOS}. 

In the light of the above results, the orthodox interpretation of Bell's argument adhered to in the critique \cite{Gill-IEEE-2} is obsolete.  

\subsection{Concerning my critique of Bell's theorem}

It turns out, however, that even the orthodox interpretation of Bell's argument harbors a serious physical mistake \cite{Oversight}. In Section~II of the critique \cite{Gill-IEEE-2}, the standard derivation of the Bell-CHSH inequality is reviewed. This derivation is well known for at least fifty years. The critique thus adds nothing new to the existing literature. Towards the end of that section the critique then quotes the following from my paper \cite{IEEE-1}: 
\begin{quote}
As innocuous as the step [that replaces the four separate averages to a single average] in the proof [of Bell-CHSH] may seem mathematically, it is, in fact, an illegitimate step physically, because what is being averaged on its right-hand are {\it unobservable} and {\it unphysical} quantities. Indeed, the pairs of measurement directions ${({\mathbf{a}},\,{\mathbf{b}})}$, ${({\mathbf{a}},\,{\mathbf{b'}})}$, ${({\mathbf{a'}},\,{\mathbf{b}})}$, and ${({\mathbf{a'}},\,{\mathbf{b'}})}$ are {\it mutually exclusive measurement directions}, corresponding to {\it incompatible} experiments which cannot be performed simultaneously.
\end{quote}
The ``step'' here concerns the assumption of the additivity of expectation values. Namely, the replacement of the sum of four separate averages with the single average of their sum:
\begin{align}
&\Bigl\langle\,{\mathscr A}_{k_1}({\mathbf{a}})\,{\mathscr B}_{k_1}({\mathbf{b}})\,\Bigr\rangle + \Bigl\langle\,{\mathscr A}_{k_2}({\mathbf{a}})\,{\mathscr B}_{k_2}({\mathbf{b'}})\,\Bigr\rangle \notag \\
&\;\;\,\;\;\;\;\;\;+\Bigl\langle\,{\mathscr A}_{k_3}({\mathbf{a'}})\,{\mathscr B}_{k_3}({\mathbf{b}})\,\Bigr\rangle - \Bigl\langle\,{\mathscr A}_{k_4}({\mathbf{a'}})\,{\mathscr B}_{k_4}({\mathbf{b'}})\,\Bigr\rangle \label{22-no} \\
&\longrightarrow\Bigl\langle{\mathscr A}_{k}({\mathbf{a}})\Big\{{\mathscr B}_{k}({\mathbf{b}})+{\mathscr B}_{k}({\mathbf{b'}})\Big\} \notag \\
&\;\;\;\;\;\;\;\;\;\;\;\;\;\;\;\;\;\;\;\;\;\;\;\;+{\mathscr A}_{k}({\mathbf{a'}})\Big\{{\mathscr B}_{k}({\mathbf{b}})-{\mathscr B}_{k}({\mathbf{b'}})\Big\}\Bigr\rangle. \label{23-n}
\end{align}
In Section~II of critique \cite{Gill-IEEE-2}, it is the following specious step,
\begin{quote}
... let us take a look at the following expression $Z:= X_1 Y_1 - X_1 Y_2 - X_2 Y_1 - X_2 Y_2$. One can rewrite it as $X_1(Y_1 - Y_2) - X_2(Y_1 + Y_2)$. ... ,
\end{quote}
that hides the assumption of the additivity of expectation values. Without this assumption Bell-CHSH inequalities cannot be derived \cite{Oversight}. I have discussed this briefly in Section~II of \cite{IEEE-3}, and extensively in \cite{Oversight}. While mathematically correct, this assumption is physically meaningless. It amounts to {\it assuming} the bound of $2$ on the CHSH correlator instead of deriving it \cite{Oversight}. Contrary to the claim made in \cite{Gill-IEEE-2}, it is an assumption over and above the assumptions of locality and realism. In fact, as noted, it is a profound physical mistake.

It is the same mistake that von~Neumann's former theorem against general hidden variable theories harbored, as I have explained in \cite{Oversight}. For observables that are not simultaneously measurable, such as the observables involved in the fermionic spin measurements in Bell-test experiments, the replacement of the sum of expectation values with the expectation value of the sum, although respected within quantum mechanics, does not hold within hidden variable theories. This was pointed out by Einstein and Grete Hermann in the 1930s within the context of von~Neumann's theorem, and again some thirty years later by Bell and others, as I have explained in \cite{Oversight}.

The problem is that, while the sum of expectation values is mathematically the same as the expectation value of the sum, as in the assumption that allows us to mathematically replace (\ref{22-no}) with (\ref{23-n}) above, and while this assumption is valid in quantum mechanics because any linear sum of operators represents another operator in Hilbert space, it is not valid for any hidden variable theory based on dispersion-free states, because the eigenvalue of a sum of operators is not the sum of individual eigenvalues (which is what ${\mathscr A}_k(\mathbf{a})$, ${\mathscr A}_k(\mathbf{a'})$, ${\mathscr B}_k(\mathbf{b})$, and ${\mathscr B}_k(\mathbf{b'})$ in fact are) when the constituent operators are non-commuting, as in the case of Bell-test experiments. This makes the replacement of (\ref{22-no}) with (\ref{23-n}) physically invalid. But without this replacement the absolute upper bound of 2 on the CHSH correlator cannot be derived. 

Once this extraordinary oversight is removed from Bell's argument and local realism is implemented correctly \cite{Oversight}, the bounds on the CHSH correlator work out to be $\pm2\sqrt{2}$ instead of $\pm2$, thereby mitigating the radical conclusions of Bell’s theorem \cite{Oversight}. Consequently, what is ruled out by the Bell-test experiments is not local realism as widely believed, but the assumption of the additivity of expectation values.

Considering the above mistake, the claim in the Introduction of the critique \cite{Gill-IEEE-2} that ``Bell's theorem has not been disproved...'' is rather ironic. For Bell's theorem has never been proven in the first place. It has only been {\it assumed}.

\subsection{Concerning the Bell inequality violations}

Much is made in the critique of the so-called new generation of loophole-free Bell-test experiments and how they exhibit violations of the Bell-CHSH inequality. By CHSH inequality the critique essentially means the inequality discovered by Boole one hundred and eleven years before Bell's famous paper of 1964 \cite{Boole-1,Boole-2}. It is a {\it mathematical} inequality involving four binary numbers: ${\mathscr A}_k(\mathbf{a})=\pm1$, ${\mathscr A}_k(\mathbf{a'})=\pm1$, ${\mathscr B}_k(\mathbf{b})=\pm1$, and ${\mathscr B}_k(\mathbf{b'})=\pm1$. As such, the absolute value of the expression (\ref{23-n}) is mathematically bounded by 2. It is {\it impossible} to violate the bound of 2 on (\ref{23-n}) with binary scalar numbers $\pm1$ that represent the results observed in the Bell-test experiments. And yet, experimenters are routinely reporting violations of the absolute bound of 2 on (\ref{23-n}) set by application of elementary arithmetic. How is that possible?

It is certainly possible to exceed the bound of 2 on (\ref{23-n}) if we can {\it switch} to four separate experiments corresponding to expression (\ref{22-no}), the absolute value of which is bounded by 4, not 2. After all, four separate experiments corresponding to expression (\ref{22-no}) is all we can hope to achieve in practice. So that is what the experimenters do. In other words, there is extraordinary bait-and-switch happening (albeit unwittingly) in every experiment that claims to have
violated the absolute bound of 2 on expression (\ref{23-n}), rationalized with statistics. 

\subsection{Concerning a denial of Bell's assumption}

In their Introductions, the critiques \cite{Gill-IEEE-1} and  \cite{Gill-IEEE-2} claim that
\begin{quote}
Bell does not take account of the geometry of space because his argument, on the side of local realism, does not depend on it in any way whatsoever.
\end{quote}
This claim, however, is not correct. I have already explained why it is not correct in Answer~2 of Appendix~B of \cite{IEEE-1} and in Subsection~IV~B of \cite{IEEE-3}. And yet, the claim has been repeated in \cite{Gill-IEEE-2} again. But nowhere in his writings has Bell stated that his theorem holds independently of the geometry of physical space. On the contrary, in Section 8 of Chapter 7 of his book \cite{Speakable}, while exploring possible strategies that can negate his theorem, he writes: ``The space time structure has been taken as given here. How then about gravitation?'' Thus Bell seems to have anticipated the use of a solution of Einstein's field equations of general relativity to overcome his theorem.

Moreover, the derivations of the Bell-CHSH inequalities explicitly require different detector settings corresponding to different directions in the physical space, which are then represented -- both theoretically and in configuring Bell-test experiments -- by ordinary vectors in ${\mathrm {I\!R}}^3$. It is unfortunate that this implicit assumption is usually not made explicit in the literature on Bell's theorem. But why must we assume such vectors to be within ${\mathrm{I\!R}^3}$? If, instead, they are embedded within quaternions constituting $S^3$ that model the physical space, then the correlations between measurement events are inevitably sinusoidal, as I have proved many times in \cite{IEEE-1,IEEE-2,Disproof,IJTP,RSOS}, in many different ways. Thus to claim that Bell's argument ``does not take account of the geometry of space'' is to reveal a {\it weakness} of Bell's argument, not its strength. 

The respective topologies of the three-dimensional spaces ${\mathrm {I\!R}}^3$ and $S^3$ are dramatically different, even though the latter differs from the former by a single point added to it at infinity:
\begin{equation}
S^3 = \,{\mathrm {I\!R}}^3 \cup \left\{\infty\right\}.
\end{equation}
While ${\mathrm {I\!R}}^3$ is an open space stretching out from the origin to infinity, $S^3$ is both a closed and compact manifold. Unlike ${\mathrm {I\!R}}^3$, the space $S^3$ has many special properties. For example, although locally it is a product space, $S^3\approx S^2\times S^1$, globally the fiber bundle of $S^3$ has no cross-section at all. Moreover, it is both a connected and simply connected manifold without boundary, so that any loop or circular path in it can be shrunk continuously to a point without leaving the manifold \cite{Bloch}.  

Now measurement results, such as the clicks of the detectors, are events in spacetime. In the EPR-Bohm or Bell-test experiments one is interested in spacelike separated coincident events in spacetime. In other words, one is interested in the spacelike separated equal-time points on a spacelike hypersurface in spacetime, such as those in ${\mathrm {I\!R}}^3$ or $S^3$. And as I have explained in \cite{IEEE-1,IEEE-2,IEEE-3}, both ${\mathrm {I\!R}}^3$ and $S^3$ are admissible spatial parts of a well known solution of Einstein's field equations of general relativity. There is thus no escape from the geometries of ${\mathrm {I\!R}}^3$ and $S^3$. But since the topological properties of ${\mathrm {I\!R}}^3$ and $S^3$ are so dramatically different, the correlations between their respective points would also be dramatically different. And yet, the critique claims that Bell's argument is insensitive to this difference. If so, then that is clearly yet another of several defects in Bell's argument.

\subsection{Concerning Equations (66) and (67) in \cite{IEEE-1}} \label{Gill-E}

Another incorrect claim in the critique \cite{Gill-IEEE-2} concerns what the measurement functions (66) and (67) defined in \cite{IEEE-1} predict: 
\begin{quote}
Christian’s explicit definition of measurement
functions results in measurement outcomes which are equal and opposite with probability one, {\it whatever the measurement settings}. [Appendix of \cite{Gill-IEEE-2}.]
\end{quote}
In its Section IV, the critique again claims that, for whatever detector settings ${\mathbf a}$  and ${\mathbf b}$ are chosen by Alice and Bob, the measurement functions (66) and (67) defined in \cite{IEEE-1} predict that the observed results ${\mathscr A}({\mathbf a},\,\lambda)$ and ${\mathscr B}({\mathbf b},\,\lambda)$ will satisfy
\begin{equation}
{\mathscr A} = - {\mathscr B} = \pm1 = \lambda, \label{mad2}
\end{equation}
with $\lambda=\pm1$ being a fair coin. But no such equation exists in \cite{IEEE-1}, or in any of my other publications \cite{IEEE-2,Disproof,IJTP,RSOS}. In fact, as I explained in Section~\ref{II} above, within the quaternionic 3-sphere model the above equation is quite meaningless. Moreover, it is very easy to verify that the measurement functions (66) and (67) defined in \cite{IEEE-1} {\it do not} predict perfect anti-correlation ${\mathscr A}{\mathscr B}=-1$ for ${\mathbf a}\not={\mathbf b}$ without violating the conservation of zero spin angular momentum. I have demonstrated this in considerable detail towards the end of Section~VIII in the original paper \cite{IEEE-1}. Thus the claim made in the critique is manifestly wrong. The above equation (\ref{mad2}) is simply made up in the critique \cite{Gill-IEEE-2}. As I have explained in Subsection~IV~E of \cite{IEEE-3} and Section~III of \cite{IEEE-2}, the perfect anti-correlation ${\mathscr A}{\mathscr B}=-1$ between the measurement results can hold within $S^3$ only for the special case when ${\mathbf a}={\mathbf b}$. In general, for ${\mathbf a}\not={\mathbf b}$, the product ${\mathscr A}{\mathscr B}$ of the observed results ${\mathscr A}$ and ${\mathscr B}$ would necessarily fluctuate between $-1$ and $+1$,
\begin{equation}
{\mathscr A}{\mathscr B}\in\{-1,\,+1\},
\end{equation}
because of the sign changes in quaternions that constitute $S^3$:
\begin{equation}
{\mathbf q}(\eta_{{\mathbf a}{\mathbf b}}+\kappa\pi,\,{\mathbf r})\,=\,(-1)^{\kappa}\,{\mathbf q}(\eta_{{\mathbf a}{\mathbf b}},\,{\mathbf r})\,\;\;\text{for}\;\,\kappa=0,1,2,3,\dots \label{spinorial}
\end{equation}
In other words, all four possible combinations of the results, ${+\,+}$, ${+\,-}$, ${-\,+}$, and ${-\,-}$, would be seen by Alice and Bob.

The violation of the conservation of spin angular momentum in the critique's claim is not difficult to detect. For ${{\mathbf a}\not={\mathbf b}}$, Eq.~(\ref{mad2}) can hold only for ${\mathbf s}_1 \not= {\mathbf s}_2$, where ${\mathbf s}_1$ and ${\mathbf s}_2$ are unit vectors about which the spin bivectors ${\mathbf L}({\mathbf s}_1,\,\lambda)$ and ${\mathbf L}({\mathbf s}_2,\,\lambda)$ are spinning after emerging from the source (cf. Fig.~\ref{Fig-2n}). This can be verified from the measurement functions (66) and (67) defined in \cite{IEEE-1}. Thus, the critique's made up equation holds in general for all choices of ${\mathbf a}$ and ${\mathbf b}$ {\it only if} the conservation of zero spin angular momentum is violated, or, equivalently, the M\"obius-like twists in the Hopf bundle of $S^3$ are ignored. That is to say, for ${{\mathbf a}\not={\mathbf b}}$, Eq.~(\ref{mad2}) holds only if the 3-sphere model is abandoned and one relapses back to the flat geometry of ${\mathrm{I\!R}^3}$, as the critiques in \cite{Gill-IEEE-1} and \cite{Gill-IEEE-2} tend to do reflexively.

I have already explained the above flaws in the critique  at great length in Subsection IV~E of my previous reply \cite{IEEE-3}, as well as in \cite{Scott}, in Appendix~C of \cite{Reply-Gill}, in Answers~9 and 14 in Appendix~B of \cite{IEEE-1}, and in Answers~6 and 7 in Appendix~B of \cite{IEEE-2}. And yet, this frequently refuted claim is repeated in \cite{Gill-IEEE-2}.

\subsection{Concerning the claim of different models}\label{E}

In its Abstract, the critique \cite{Gill-IEEE-2} claims that my paper \cite{IEEE-1} ``... contains several conflicting models.'' And again in its Sections~III and IV it refers to the 3-sphere model presented in \cite{IEEE-1} as my ``first model'' and ``second model'', implying that I have proposed two different models. But throughout my work published in  
\cite{IEEE-1,IEEE-2,Disproof,IJTP,RSOS}, and especially in my paper \cite{IEEE-1}, I have proposed only {\it one} model for the singlet correlations, and that model is the quaternionic 3-sphere model summarized in Section III of \cite{IEEE-3}. This suggests that the critique \cite{Gill-IEEE-2} is based on quite a mistaken understanding of what is presented in \cite{IEEE-1}.

In reality, what is presented in Sections IV and VII of \cite{IEEE-1} are not different models but two different {\it representations} of one and the same quaternionic 3-sphere model. And there is no ``conflict'' of any kind between these two representations. They complement each other perfectly, and prove that, no matter which representation of the quaternionic 3-sphere is used to perform calculations,  the correlations predicted by the quantum mechanical singlet state can be understood as local, realistic, and deterministic correlations among the limiting scalar points ${\mathscr A}=\pm1$ and ${\mathscr B}=\pm1$ of the 3-sphere, provided it is taken to model the three-dimensional physical space in which we are confined to perform our Bell-test experiments. 

It is also important to note that no argument or proof is provided in \cite{Gill-IEEE-2} (or anywhere else) in support of its incorrect claim. The critique alleges ``conflicting models'', but does not provide any evidence or demonstration of the alleged conflict. 

It is well known that mathematically a 3-sphere can be represented in several different ways. In Sections IV and V of my paper \cite{IEEE-1} it is represented {\it without} using the full arsenal of Geometric Algebra in a manner that brings out an unforeseen link between the quaternionic 3-sphere model and Pearle's local hidden variable model for the singlet correlations \cite{Pearle}. On the other hand, in Sections VII, VIII, and IX of \cite{IEEE-1} the 3-sphere is represented using the elegant and powerful language of Geometric Algebra. It is the strength of my analysis that two entirely different representations of the 3-sphere are shown to reproduce the same singlet correlations predicted by quantum mechanics. What is more, pedagogically each representation illuminates the quaternionic 3-sphere model in a different way. The critique \cite{Gill-IEEE-2}, however, claims the two representations of the model to be two different models. That is analogous to claiming that Heisenberg and Schr\"odinger representations of quantum theory are two different theories.

\subsection{Concerning Pearle's calculation within $\boldsymbol{S^3}$}\label{F}

In its Abstract, critique \cite{Gill-IEEE-2} claims that ``Most of [my] paper is devoted to a model based on the detection loophole due to Pearle (1970).'' But only a few formulae from Pearle's paper are used in only two out of twelve sections and two appendices of my paper \cite{IEEE-1}, and neither of the two sections are exclusively concerned with his detection loophole analysis. This suggests that the critique's reading of \cite{IEEE-1} is mistaken.

In fact, far from being ``devoted'' to Pearle's detection loophole model \cite{Pearle}, the discussion in Section V of \cite{IEEE-1} is about what has been {\it missed} in Pearle's analysis in \cite{Pearle}, and how the singlet correlation can be recovered once that defect in \cite{Pearle} is rectified. The state space for the singlet spin system used by Pearle is a well known representation of the rotation group SO(3), viewed as a unit ball within ${\mathrm{I\!R}^3}$, with each point of it representing a rotation about a vector ${\mathbf r}$ of length $0\leqslant r \leqslant 1$ from the origin, and the product ${\pi r}$ representing a rotation by angle $\psi=\pi r$. By contrast, the state space used in the quaternionic 3-sphere model of the singlet correlations presented in \cite{IEEE-1} uses a representation of the double covering group of SO(3). Namely, the group ${\mathrm{SU(2)}}$, which is homeomorphic to ${S^3}$, and can be constructed by identifying the boundaries of two SO(3) balls, providing the double covering of the group SO(3). The state space ${\mathrm{SU}(2)\cong S^3}$ is thus topologically nontrivial, and sensitive to the spinorial sign changes in the quaternions that constitute $S^3$. By contrast, Pearle's sate space SO(3) is insensitive to such sign changes. Thus, what is overlooked in Pearle's analysis is the spinorial sign changes in the quaternions described in Eq.~(\ref{spinorial}) above. 

\begin{figure}[t]
\centering
\includegraphics[scale=0.5]{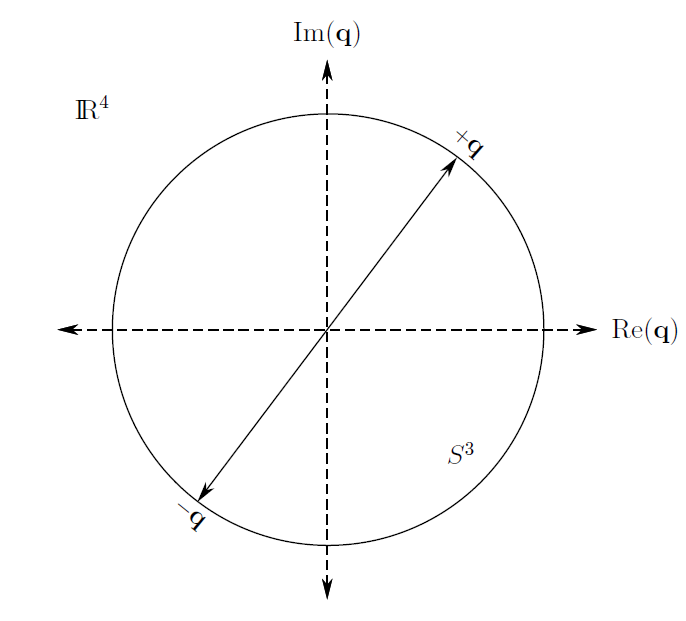}
\caption{The configuration space SO(3) of all possible rotations in physical space is obtained by identifying the antipodal points of $S^3$ --- {\it i.e.}, by identifying every quaternion ${+{\mathbf q}}$ in ${S^3}$ with its antipodal quaternion ${-{\mathbf q}}$ in ${S^3}$. After \cite{IJTP}.}
\label{Fig-7}
\end{figure}

To appreciate this better, recall that, in general, the group ${\mathrm{SU}(2)\cong S^3}$ represents rotations of objects relative to other objects. Such rotations are sensitive to spinorial sign changes and thus fermionic  \cite{IJTP}. On the other hand, local observations of rotations that do not refer to external objects are not sensitive to sign changes in the quaternions that constitute the group ${\mathrm{SU}(2)\cong S^3}$. In other words, the quaternions ${-\,{\mathbf q}(\psi,\,{\mathbf r})}$ and ${+\,{\mathbf q}(\psi,\,{\mathbf r})}$ in $S^3$ represent one and the same rotation in the physical space ${\mathrm{I\!R}^3}$. There are thus twice as many elements in the set ${S^3}$ of all unit quaternions than there are points in the configuration space SO(3) of all possible rotations in the physical space. This is because every pair of quaternions constituting the antipodal points of ${S^3}$ represent one and the same rotation in ${\mathrm{I\!R}^3}$ (cf. Fig.~\ref{Fig-7}). This can be verified by recalling how a quaternion and its antipode can rotate a bivector ${{\mathbf J}({\mathbf r})}$ about ${\mathbf r}$ to a bivector ${{\mathbf J}({\mathbf r'})}$ about ${\mathbf r'}$:
\begin{equation}
{\mathbf J}({\mathbf r'})\,=\,(+\,{\mathbf q})\,{\mathbf J}({\mathbf r})\,(+\,{\mathbf q})^{\dagger}
\,=\,(-\,{\mathbf q})\,{\mathbf J}({\mathbf r})\,(-\,{\mathbf q})^{\dagger}. \label{rot}
\end{equation}
Thus ${S^3}$, or more precisely the group SU(2) homeomorphic to $S^3$, represents a universal double covering of the rotation group SO(3) so that the quaternions ${-\,{\mathbf q}(\psi,\,{\mathbf r})}$ and ${+\,{\mathbf q}(\psi,\,{\mathbf r})}$ represent one and the same rotation in the physical space ${\mathrm{I\!R}^3}$. SO(3) is thus a group of observable rotations in physical space and can be constructed by identifying each quaternion ${-\,{\mathbf q}(\psi,\,{\mathbf r})}$ in ${S^3}$ with its antipodal quaternion ${+\,{\mathbf q}(\psi,\,{\mathbf r})}$ in ${S^3\hookrightarrow \mathrm{I\!R}^4}$. The space that results is the real projective space
\begin{equation}
{\mathrm{I\!R}}{\mathrm P}^3\cong{\mathrm{SO}}(3)\approx
\left({\mathrm{SU}}(2)\cong S^3\right)/\{-1,\,+1\},
\end{equation}
which is simply the set of all lines through the origin of ${{\mathrm{I\!R}}^4}$.

\begin{figure}[t]
\centering
\includegraphics[scale=0.43]{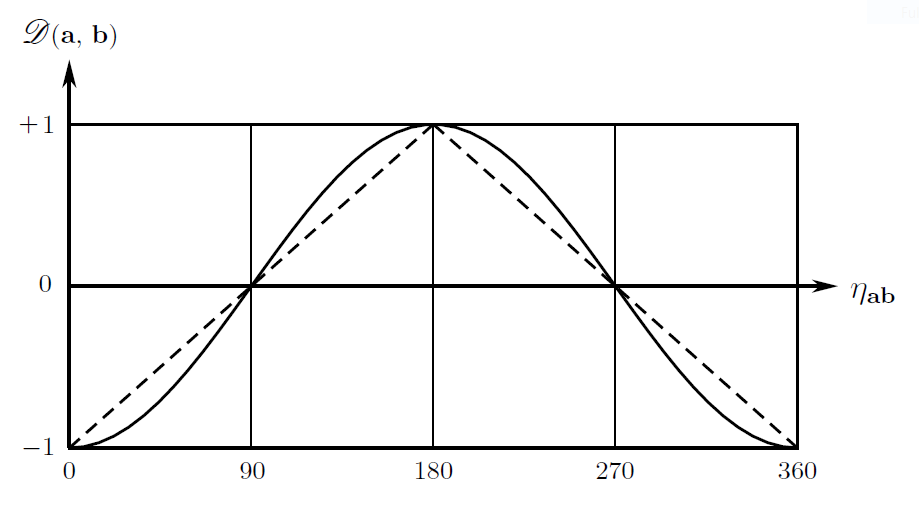}
\caption{Comparison of the geodesic distances ${{\mathscr D}({\mathbf a},{\mathbf b})}$ on ${\mathrm{SU}(2)\cong S^3}$ and ${\mathrm{SO}(3)\cong {\mathrm{I\!R}}{\mathrm P}^3}$ as functions of a half of the rotation angle. The dashed lines depict the geodesic distances on ${\mathrm{SO}(3)\cong {\mathrm{I\!R}}{\mathrm P}^3}$, which was taken to be the state space of the singlet system by Pearle in \cite{Pearle}, whereas the solid curve depicts their horizontal lift to the covering group ${\mathrm{SU}(2)\cong S^3}$, which is taken to model the three-dimensional physical space in the quaternionic 3-sphere model of the singlet correlations I have proposed in \cite{IEEE-1,IEEE-2,Disproof,IJTP,RSOS}. After \cite{IJTP}.}
\label{Fig-2}
\end{figure}

There are thus two preimages in ${S^3}$, namely ${+\,{\mathbf q}}$ and ${-\,{\mathbf q}}$ for each rotation within SO(3). In other words, the projective space ${{\mathrm{I\!R}}{\mathrm P}^3}$ is the quotient of ${S^3}$ by the map ${{\mathbf q}\mapsto -\,{\mathbf q}}$, which can be expressed as ${\varphi:S^3\rightarrow{\mathrm{I\!R}}{\mathrm P}^3}$. This quotient map renders the topologies of the spaces ${S^3}$ and ${{\mathrm{I\!R}}{\mathrm P}^3}$ quite distinct from one another. For instance, the space ${S^3}$ turns out to be simply-connected, whereas the space ${{\mathrm{I\!R}}{\mathrm P}^3}$ is connected but not simply-connected. Consequently, the geodesic distances ${{\mathscr D}({\mathbf a},\,{\mathbf b})}$ between two quaternions ${{\mathbf q}(\psi_{\mathbf a},\,{\mathbf a})}$ and ${{\mathbf q}(\psi_{\mathbf b},\,{\mathbf b})}$ representing two different rotations within ${\mathrm{I\!R}^3}$ can be measurably different in the manifolds ${\mathrm{SU}(2)\cong S^3}$ and ${\mathrm{SO}(3)\cong \mathrm{I\!R}{\mathrm P}^3}$, providing a signature of spinorial sign changes between ${{\mathbf q}(\psi_{\mathbf a},\,{\mathbf a})}$ and ${{\mathbf q}(\psi_{\mathbf b},\,{\mathbf b})}$. This signature is shown in Fig.~\ref{Fig-2}, and discussed in considerable detail in~\cite{IJTP}.

The failure of Pearle's theoretical model in \cite{Pearle} to take into account the spinorial sign changes captured in (\ref{spinorial}) is rectified in my paper \cite{IEEE-1} with amicable consequences. If we include the spinorial sign changes in Pearle's analysis, then the relation between the rotation angle ${{\pi}r}$ in Pearle's state space ${\mathrm{SO}(3)\cong {\mathrm{I\!R}}{\mathrm P}^3}$ and the rotation angle ${2\eta_{{\mathbf z}{\mathbf s}_o}}$ in the state space ${{\mathrm{SU}(2)}\cong S^3}$ I have used in \cite{IEEE-1,IEEE-2,Disproof,IJTP,RSOS} works out to be
\begin{numcases}
{\cos\left(\frac{\pi}{2}r\right)=}
-1+\frac{2}{\sqrt{1+3\left(\frac{\eta_{{\mathbf z}{\mathbf s}_o}}{\kappa\pi}\right)}\,}= f(\eta_{{\mathbf z}{\mathbf s}_o}), &  \label{f} \\
-1+\frac{2}{\sqrt{4-3\left(\frac{\eta_{{\mathbf z}{\mathbf s}_o}}{\kappa\pi}\right)}\,}=f(\pi-\eta_{{\mathbf z}{\mathbf s}_o}). & \label{fp}
\end{numcases}
There is then a one-to-one correspondence between the initial spin states emitted by the source and the final spin states detected by the remote detectors of Alice and Bob. Thus, {\it every initial state that is emitted by the source is detected by the detectors, and vice versa}. Consequently, unlike Pearle's model, the model presented in \cite{IEEE-1} is not about data rejection or detection loophole. In the latter model, the fraction ${{\mathrm g}(\eta_{{\mathbf a}{\mathbf b}})}$ of events in which both particles are detected is exactly one:
\begin{equation}
{\mathrm g}(\eta_{{\mathbf a}{\mathbf b}})\!=\frac{P_{12}^{+-}(\eta_{{\mathbf a}{\mathbf b}})}{\frac{1}{2}\cos^2\!\left(\frac{\eta_{{\mathbf a}{\mathbf b}}}{2}\right)}=\frac{P_{12}^{++}(\eta_{{\mathbf a}{\mathbf b}})}{\frac{1}{2}\sin^2\!\left(\frac{\eta_{{\mathbf a}{\mathbf b}}}{2}\right)}=1\;\forall\;\eta_{{\mathbf a}{\mathbf b}}\in[0,\,\pi]. \label{cl48}
\end{equation}
This is made very clear in \cite{IEEE-1}, in the following passage:
\begin{quote}
Clearly, a measurement event cannot occur if there does not exist a state which can bring about that event. Since the initial state of the system is specified by the pair ${({\mathbf e}_o,\,{\mathbf s}_o)}$ and not just by the vector ${{\mathbf e}_o}$, there are no states of the system for which ${|\cos(\,\eta_{{\mathbf n}{\mathbf e}_o})|\,<\,f(\eta_{{\mathbf z}{\mathbf s}_o})}$ for {\it any} vector ${\mathbf n}$. Thus a measurement event cannot occur for ${|\cos(\,\eta_{{\mathbf n}{\mathbf e}_o})|\,<\,f(\eta_{{\mathbf z}{\mathbf s}_o})}$, no matter what ${\mathbf n}$ is. As a result, there is a one-to-one correspondence between the initial state ${({\mathbf e}_o,\,{\mathbf s}_o)}$ selected from the set (31) and the measurement events ${\mathscr A}$ and ${\mathscr B}$ specified by the Eqs.~(34) and (35). This means, in particular, that the ``fraction'' ${{\mathrm g}(\eta_{{\mathbf a}{\mathbf b}})}$ in our model is equal to 1 for all ${\eta_{{\mathbf a}{\mathbf b}}}$, dictating the vanishing of the probabilities
\begin{equation}
P_{12}^{00}(\eta_{{\mathbf a}{\mathbf b}})=1+{\mathrm g}(\eta_{{\mathbf a}{\mathbf b}})-2\,{\mathrm g}(0)=0, \tag{49}
\end{equation}
which follows from Pearle's Eq.~(9). Moreover, from his Eq.~(8) we also have
\begin{equation}
{P_{12}^{+0}(\eta_{{\mathbf a}{\mathbf b}})=\frac{1}{2}\left[\,{\mathrm g}(0)-{\mathrm g}(\eta_{{\mathbf a}{\mathbf b}})\right]}, \notag
\end{equation}
giving
\begin{align}
P_{12}^{+0}(\eta_{{\mathbf a}{\mathbf b}})=P_{12}^{-0}(\eta_{{\mathbf a}{\mathbf b}})
&=P_{12}^{0+}(\eta_{{\mathbf a}{\mathbf b}}) \notag \\
&=P_{12}^{0-}(\eta_{{\mathbf a}{\mathbf b}})=0. \tag{50}
\end{align}
Together with the probabilities for individual detections,
\begin{align}
P_1^{+}({\mathbf a})=P_1^{-}({\mathbf a})
&=P_2^{+}({\mathbf b}) \notag \\
&=P_2^{-}({\mathbf b})=\frac{1}{2}\,{\mathrm g}(0)=\frac{1}{2}\,, \tag{51}
\end{align}
the correlation between ${\mathscr A}$ and ${\mathscr B}$ then works out to be
\begin{align}
{\cal E}&({\mathbf a},\,{\mathbf b}) \notag \\
&=\lim_{\,n\,\gg\,1}\left[\frac{1}{n}\sum_{i\,=\,1}^{n}\,
{\mathscr A}({\mathbf a};\,{\mathbf e}^i_o,\,{\mathbf s}_o^i)\;{\mathscr B}({\mathbf b};\,{\mathbf e}^i_o,\,{\mathbf s}_o^i)\right] \notag \\
&=\,\frac{P_{12}^{++}\,+\,P_{12}^{--}\,-\,P_{12}^{+-}\,-\,P_{12}^{-+}}{P_{12}^{++}\,+\,P_{12}^{--}\,+\,P_{12}^{+-}\,+\,P_{12}^{-+}}
\notag \\
&=\,-\,\cos\left(\eta_{{\mathbf a}{\mathbf b}}\right). \label{corsum} \tag{52}
\end{align}
Since all of the probabilities predicted by our local model in ${S^3}$ match exactly with the corresponding predictions of quantum mechanics, the violations of not only the CHSH inequality, but also Clauser-Horne inequality follow.
\end{quote}
The critique \cite{Gill-IEEE-2}, on the other hand, does not mention any of the foregone reasoning and analysis. That is unfortunate, for it is very clear from this analysis that the usual interpretation of Pearle's model as a detection loophole model is mistaken.

\subsection{Concerning a computer code included in \cite{IEEE-1}}

By now it should be evident that the critique in \cite{Gill-IEEE-2} has not understood the 3-sphere model I have presented in \cite{IEEE-1}. It is therefore not surprising that it has also misunderstood one of the six numerical simulations of the singlet correlations discussed in my paper. By relapsing back to the flat geometry of ${\mathrm{I\!R}^3}$, the critique claims that the code of the simulation cited as [40] in \cite{IEEE-1} is a simulation of Pearle's ``data rejection'' model. But in the previous subsection we established that interpreting the theoretical model proposed by Pearle in \cite{Pearle} as a data rejection or detection loophole model is a mistake. 

To begin with, I have discussed, not one, but six different numerical simulations of the singlet correlations in my paper, cited as references [40], [41], [45], [46], [47], and [48]. The critique, however, discusses only one of them, namely, the one cited as reference [40] in \cite{IEEE-1}. More importantly, as I discussed in Subsection~\ref{E} above, it is quite well known that, mathematically, a 3-sphere can be represented in several different ways. In Sections IV, V, and VI of my paper \cite{IEEE-1} the 3-sphere is represented without using   Geometric Algebra in such a way that it almost aligns with Pearle's hidden variable model for the singlet correlations \cite{Pearle}. This makes it possible to parallel the detection loophole type simulation analysis but without failing to detect any initial states emerging form the source in the EPR-Bohm type experiment. On the other hand, in Sections VII, VIII, and IX of \cite{IEEE-1} the same 3-sphere is represented using the powerful and elegant language of Geometric Algebra, which facilitates the use of a {\it GAViewer} program for simulating the resulting singlet correlations. It is the strength of my analysis in \cite{IEEE-1} that two entirely different representations of the 3-sphere are shown to reproduce {\it the same} singlet correlations predicted by quantum mechanics.

Now for the Geometric Algebra based simulations [45] to [48] cited in \cite{IEEE-1} the issue of data rejection does not arise, and the critique in \cite{Gill-IEEE-2} does not claim that it does. On the other hand, the simulations cited in \cite{IEEE-1} as [40] and [41] simulates the singlet correlations using the first of the two representations of the 3-sphere mentioned in the previous paragraph. For them, the critique in \cite{Gill-IEEE-2} claims that the simulation cited as [40] in \cite{IEEE-1} depends on ``the post-selection of data'' and thus uses ``data rejection.'' But this claim is not correct, as can be easily verified by examining the simulation in the context of the 3-sphere model discussed in the Sections IV, V, and VI of \cite{IEEE-1}. {\it There is no post-selection of data in this simulation}. There is only pre-selection of only those initial states that can exist within the 3-sphere. Those are the only initial states that are meaningful in the model, and none of them are rejected. Consequently, as noted in the Subsection~\ref{F} above, there is one-to-one correspondence between the initial states of the system and those states that are detected by Alice and Bob. 

This fact is more evident in the simulation cited as [41] in \cite{IEEE-1}, but the critique in \cite{Gill-IEEE-2} has overlooked the analysis in [41]. In that simulation it is demonstrated how, by setting the distribution function f = 0, the geometry of $S^3$ reduces to that of the flat geometry of ${\mathrm{I\!R}^3}$, and that, in turn, reduces the sinusoidal correlations to the saw-tooth shaped correlations.

There is another simulation that is worth mentioning here, even though I have neglected citing it in \cite{IEEE-1}; namely, Ref.~\cite{Simulation-C} below. The analysis in this simulation is aimed at countering the incorrect claims of ``post-selection of data”, ``use of data rejection'', or ``exploiting the detection loophole'' leveled in the critique \cite{Gill-IEEE-2} against the simulation cited as [40] in \cite{IEEE-1}. From the detailed numerical analysis presented in the simulation \cite{Simulation-C} cited below it is abundantly clear that any claim of data rejection is nothing more than wishful thinking.

\subsection{Concerning Lasenby's critique of my work}

Next, let me briefly comment on the critique \cite{Gill-IEEE-2}'s appeal to Lasenby's critique \cite{Lasenby-AACA} in support of its own mistaken claims:
\begin{quote}
... Lasenby (2020) \cite{Lasenby-AACA} shows that a central and purely algebraic result in Christian (2018) \cite{RSOS} (published in Royal Society Open Science) is wrong. ... Lasenby identifies exactly the same GA errors as I did in my papers \cite{Gill-IEEE-1}, \cite{Gill-Entropy} ...
\end{quote}
Lasenby does make such claims in \cite{Lasenby-AACA}. Unfortunately, while his critique in \cite{Lasenby-AACA} is more cogently presented compared to those in \cite{Gill-IEEE-1,Gill-IEEE-2}, and \cite{Gill-Entropy}, it too succumbs to some of the same conceptual and mathematical mistakes I have brought out from \cite{Gill-IEEE-1,Gill-IEEE-2,Gill-Entropy} in \cite{IEEE-3,Reply-Gill,Reply-Gill-IJTP,Scott}, Appendix~B of \cite{IEEE-1}, and Appendix~B of \cite{IEEE-2}. This is not surprising because they are uncritically borrowed from \cite{Gill-IEEE-1} and  \cite{Gill-Entropy}, as acknowledged in \cite{Lasenby-AACA}. In particular, the claims made in \cite{Lasenby-AACA} are not correct. I have demonstrated them to be incorrect in \cite{Reply-Lasenby} and \cite{Reply-to-Lasenby}.

\subsection{Concerning the alleged survey of my work}

In its Introduction, critique \cite{Gill-IEEE-2} claims to have surveyed my work on Bell's theorem in the critique \cite{Gill-Entropy} (see also \cite{Gill-apology}):
\begin{quote}
The present author surveyed Christian's work on Bell's theorem from 2007 to 2019 in \cite{Gill-Entropy}, ... Soon after that, I published a ``Comment'' \cite{Gill-IEEE-1} to Christian's (2020) companion {\it IEEE Access} paper \cite{IEEE-2}.
\end{quote}
Unfortunately, the alleged survey of my work in \cite{Gill-Entropy} is anything but a survey of my work. It contains a large number of incorrect assertions, made up history, and armchair sociology and psychology, that even a minimal fact checking would reveal to be unfit for any scientific discourse. It also contains numerous mathematical and conceptual mistakes, some of which I have brought out in \cite{IEEE-3,Reply-Gill,Reply-Gill-IJTP,Scott}. The mistakes in \cite{Gill-Entropy} begin with its very title, which reads: “Does geometric algebra provide a loophole to Bell’s theorem?” But my work in \cite{IEEE-1,IEEE-2,Disproof,IJTP,RSOS} on quantum correlations has nothing to do with any loopholes. Thus, the title reveals a lack of understanding of what my work is all about. The other mistakes in \cite{Gill-Entropy,Gill-apology} are too numerous to bring out here, but some of them have been published in the reviewer reports \cite{Gill-reports}. Also, the entire presentation in \cite{Gill-Entropy} is marred by the use of matrices and vector "algebra" (which, of course, does not form an algebra), thus missing the very point of my use of Geometric Algebra in \cite{IEEE-1,IEEE-2,Disproof,IJTP,RSOS}. This is quite a serious conceptual mistake in \cite{Gill-Entropy}.

\section{Conclusion}

The common defect in the critiques \cite{Gill-IEEE-1}, \cite{Gill-IEEE-2}, and \cite{Gill-Entropy} is that, instead of engaging with the original quaternionic 3-sphere model presented in my papers \cite{IEEE-1,IEEE-2,Disproof,IJTP,RSOS,IEEE-3} using Geometric Algebra, they insist on criticizing entirely unrelated flat space models based on matrices and vector "algebra." This logical fallacy by itself renders the critiques invalid. Nevertheless, in this paper I have addressed every claim made in the critique \cite{Gill-IEEE-2} and the critiques it relies on, and demonstrated, point by point, that none of the claims made in the critiques are correct. I have demonstrated that the claims made in the critique \cite{Gill-IEEE-2} are neither proven nor justified. In particular, I have demonstrated that, contrary to its claims, critique \cite{Gill-IEEE-2} has not found any mistakes in my paper \cite{IEEE-1}, or in my other related papers, either in the analytical model for the singlet correlations or in its event-by-event numerical simulations. Moreover, I have brought out a large number of mistakes and incorrect statements from the critique \cite{Gill-IEEE-2} and the critiques it relies on. Some of these mistakes are surprisingly elementary.

In its Conclusion, the critique \cite{Gill-IEEE-2} claims that my ``paper \cite{IEEE-1} makes no contribution to the ongoing debates concerning Bell's theorem.'' However, contributing to the ongoing debates concerning Bell's theorem has never been my goal. In my view, Bell's theorem is a deeply flawed argument, and therefore it is irrelevant for the future of physics. With the problem of reconciling general relativity with quantum mechanics on the back of my mind \cite{Newton-Cartan}, my goal has always been understanding the origins and strengths of quantum correlations local-realistically, in terms of the geometry and topology of spacetime, as envisaged by Einstein \cite{Pais}. In my view, I have taken the first steps in that direction in \cite{IEEE-1,IEEE-2,Disproof,IJTP,RSOS}. Nothing the critique \cite{Gill-IEEE-2} and the critiques it relies on have claimed undermines this work. In \cite{IEEE-1,IEEE-2,Disproof,IJTP,RSOS} I have shown that the {\it raison d'\^etre} of the observed strong correlations is not quantum entanglement {\it per se} (which is merely a placeholder for the strong correlations), but the geometry and topology of the physical space in which we are confined to perform our experiments. I have demonstrated in \cite{IEEE-1,IEEE-2,Disproof,IJTP,RSOS} that the origins and strengths of strong correlations can be easily understood local-realistically if we model the physical space as $S^3$ instead of ${\mathrm{I\!R}^3}$. There is no need to succumb to nonlocality, superdeterminism, retrocausality, or conspiracy.

\end{document}